\documentclass[aps,groupedaddress,showpacs,showkeys,nofootinbib]{revtex4}
\usepackage{graphicx,amssymb,amsmath,bm}

\newcommand{\be}{\begin{equation}}
\newcommand{\ee}{\end{equation}}
\newcommand{\bea}{\begin{eqnarray}}
\newcommand{\eea}{\end{eqnarray}}
\newcommand{\nn}{\nonumber}

\def\de#1/de#2{\frac{\partial {#1}}{\partial {#2}}}
\begin{document}
\title{Exact solutions for Weyl fermions with gravity}
\author{ Roberto Cianci\footnote{E-mail: cianci@dime.unige.it},
Luca Fabbri\footnote{E-mail: fabbri@dime.unige.it},
Stefano Vignolo\footnote{E-mail: vignolo@dime.unige.it}}
\affiliation{DIME Sez. Metodi e Modelli Matematici, Universit\`{a} di Genova\\
Piazzale Kennedy, Pad. D - 16129 Genova, Italia}
\date{\today}
\begin{abstract}
We consider the single-handed spinor field in interaction with its own gravitational field described by the set of field equations given by Weyl field equations written in terms of derivatives that are covariant with respect to the gravitational connection plus Einstein field equations soured with the energy tensor of the spinor: for the Weyl spinor and the ensuing spacetime of Weyl-Lewis-Papapetrou structure, we will find all exact solutions. The obtained solution for the metric tensor is that of a PP-wave spacetime while the spinor field is a flag-dipole.
\end{abstract}
\maketitle
\section{Introduction}
Not long after Einstein wrote the field equations of gravitation, Schwarzschild applied them outside a stationary and spherically symmetric matter distribution obtaining the corresponding exact solution that describes the surrounding gravitational field; soon after Maxwell wrote the field equations of electrodynamics, Maxwell himself applied them in the vacuum finding exact solutions describing the propagation of electromagnetic waves; and it took even shorter for Dirac to write the spinorial field equations finding in the free case the exact solutions representing the material wave function. All these instances refer to cases of paramount importance in physics because the knowledge of those exact solutions provided a detailed description of a given physical system, whether it was planetary precession and light bending, light-rays propagation, or the properties of electrons; on the other hand, such solutions are obtained in the vacuum or for free fields, that is for isolated systems. Interacting systems can of course be studied, but in that case the quest to find the exact solutions immediately proves to be a much tougher task to accomplish.

A first example of such a kind of interacting system can be found again when Einstein equations are sourced by the energy density of an electrodynamic field whose Maxwell equations are written with the covariant derivative containing the spacetime connection, and for this system it was Kerr who studied the axially symmetric case finding exact solutions; to our knowledge however, this first example is also the last. And to an attentive examination, this is not a fully coupled system either, because if the gravitational and electrodynamic fields are coupled, nevertheless the material charged distribution is thought to be circumscribed within the singularity of the Kerr metric and it does not appear as the current density of the electrodynamic field equations nor as the energy density of the gravitational field equations. Interacting systems of matter and its own interactions can be discussed, but in this case the search for exact solutions has clearly become harder and harder to achieve.

If one wishes to study the complete system of Einstein equations sourced by the energy density of electrodynamic and material fields, with Maxwell equations having the covariant derivatives containing the spacetime connection and sourced by the current density of the matter field, and the matter field equations having all covariant derivatives containing both the spacetime and the gauge connection, then the system becomes considerably more difficult, and in fact no exact solution has been found yet.

On the other hand, looking for exact solutions of the complete and fully coupled system of matter and its own interactions is too fundamental a problem to be abandoned because of its difficulty. So if one were to embark on the enterprise of looking for such exact solutions, the search may start by considering the simplest system of all.

The simplest system one may think to study would have to consider the spinor field in interaction with its own gravitational field because no field can exist without an energy density; however, considering spinors without interactions of electrodynamic essence is a way to simplify the problem since we do know spinors that have no electric charge: therefore, a first assumption we will consider is that of focusing on neutral spinor fields.

Another possible simplification may come from the fact that spinors, despite being fundamental, nevertheless are not irreducible objects, but they are constituted by two chiral parts which, despite being in interaction, nevertheless are independent; however, their interaction is encoded in the presence of the mass term and as a consequence the assumption of masslessness would allow for the complete splitting of left-handed and right-handed projections, and the possibility to study them separately: hence, a second assumption will be that of masslessness and the consequent selection of a single-handed component for the spinor field.

In recent times, extensions of the standard model of particle physics have had to deal with the inclusion of a possible mechanism of neutrino oscillations, and the subsequent connection to neutrino masses, but there is a problem in such a scenario: although it has been demonstrated that neutrino masses would imply neutrino oscillations, the converse has never been proven; as such, we might be open to the possibility that neutrinos oscillate even if massless because of the action of some other mechanism. This possibility is made even more cogent by the fact that apart from the phenomenon of oscillation there is no other theoretical reason for neutrinos to be massive; further still, no experimental evidence of neutrino masses is present. If this scenario were true, then neutrinos would be massless, and they would constitute the physical prototype of the exact solutions we are searching in this paper.

If this scenario were not true, and neutrinos were massive after all, then there would be no particle that can be modelled by a possible exact solution at the moment; but there could have been in the earliest stages of cosmic evolution. In fact, before cosmic inflation, the universe was supposed to be in a configuration of unbroken electroweak symmetry, where all particles were massless. Electroweak symmetry would impose an $SU(2)\times U(1)$ gauge interaction but still there may be situations in which the spacetime curvature would be the dominant contribution, and there the particle would be modelled by the exact solution we plan to find.

And even if there were no situation in which the spacetime curvature dominate, an exact solution for a single-handed spinor may still be of mathematical interest because it could be taken as the least-order contribution of the small-mass perturbative solution for a massive spinor field. That is, the solution for the massive spinor field could be obtained as a series expansion in the mass, with the exact solution in the massless case as the leading order, and solving the Dirac equation at the first order in the small mass term.

As it stands, there may be several applications for a solution that, being exact, would be important also in itself.

In this paper, we will look for such exact solution.
\section{Geometric and material field equations}
To begin, we start with the fundamental kinematic definitions and the form of the dynamical action we will employ.

Throughout the paper, partial derivatives of a given function $f(x^h)$ are denoted by $f_{x^h}:=\frac{\partial f}{\partial x^h}$. Latin and Greek indices run from $1$ to $4$. The metric tensor of the spacetime is denoted by $g_{ij}$ while the tetrad field associated with a given metric is indicated by $e^\mu_i$ in such a way that $\eta^{\mu\nu}=g_{ij}e^i_\mu e^j_\nu$ with $\eta^{\mu\nu}=\mathrm{diag}(-1,-1,-1,1)$ being the Minkowski metric, and with inverse $e^i_\mu$ verifying therefore $e^i_\mu e^\mu_j=\delta^i_j$ and $e^\mu_i e^i_\nu= \delta^\mu_\nu$ in terms of Kronecker delta. Dirac matrices are indicated by $\gamma^\mu$ and $\Gamma^i := e^i_\mu\gamma^\mu$ with $\gamma^5 =i\gamma^4\gamma^1\gamma^2\gamma^3$ and chiral representation is used. The spinorial-covariant derivatives of a Dirac field $\psi$ are expressed as 
\begin{equation}\label{defdsm}
D_i\psi=\psi_{x^i}-\Omega_{{i}}\psi
\end{equation}
where the spinorial-connection $\Omega_i$ is given by
\begin{equation}\label{1.1}
\Omega_i := - \frac{1}{4}g_{jh}\omega_{i\;\;\;\nu}^{\;\;\mu}e_\mu^j e^\nu_k\Gamma^h\Gamma^k
\end{equation}
with $\omega_{i\;\;\;\nu}^{\;\;\mu}$ being the coefficients of the spin-connection associated through the relation
\begin{equation}\label{1.2}
\Gamma_{ij}^{\;\;\;h} = \omega_{i\;\;\;\nu}^{\;\;\mu}e_\mu^h e^\nu_j + e^{h}_{\mu}\partial_{i}e^{\mu}_{j}
\end{equation}
to the linear connection $\Gamma_{ik}^{\;\;\;j}$ as usual. Natural units $\hbar=c=8\pi G=1$ are used.

We will consider a Dirac field coupled to gravity in the Einstein-Sciama-Kibble theory. The Dirac Lagrangian is
\begin{equation}\label{1.1bis}
L_\mathrm{D} =\frac{i}{2}\left( \bar{\psi}\Gamma^iD_i\psi-D_i\bar{\psi}\Gamma^i\psi\right)-m\bar{\psi}\psi
\end{equation}
where $m$ is the mass of the spinor: the total Lagrangian can be varied with respect to all fields yielding the Einstein field equations coupling the Ricci curvature $R_{ij}$ to the energy density
\begin{equation}\label{1.3a}
R_{ij} -\frac{1}{2}Rg_{ij}= \Sigma_{ij} 
\end{equation}
the Sciama-Kibble field equations coupling the Cartan torsion $T_{ij}^{\;\;\;h}$ to the spin density
\begin{equation}\label{1.3b}
T_{ij}^{\;\;\;h} = S^{\;\;\;h}_{ij}
\end{equation}
where
\begin{equation}\label{1.4}
\Sigma_{ij} := \frac{i}{4} \left( \bar\psi\Gamma_{i}{D}_{j}\psi - {D}_{j}\bar{\psi}\Gamma_{i}\psi \right) -\frac{1}{2}L_\mathrm{D}\,g_{ij}
\end{equation}
and
\begin{equation}\label{1.5}
S_{ij}^{\;\;\;h}:=\frac{i}{16}\bar\psi\left\{\Gamma^{h},[\Gamma_i,\Gamma_j]\right\}\psi
\end{equation}
are respectively the energy and the spin density tensors and
\begin{equation}\label{1.3c}
i\Gamma^{h}D_{h}\psi + \frac{i}{2}T_h\Gamma^h\psi- m\psi=0
\end{equation}
are the Dirac field equations determining the dynamical behaviour of the spinor field.

After separating all curvatures and covariant derivatives into purely metric curvatures and covariant plus torsional terms, using the torsion-spin coupling equations \eqref{1.3b} it is possible to substitute all torsion contributions into spin-spin contact potentials \cite{FV,VFC}: the significant part of the Einstein field equations results to be the symmetric one, which can be written in the final form as a relationship linking the purely metric Ricci curvature tensor and scalar $\tilde{R}_{ij}$ and $\tilde R$ to the purely metric energy density tensor as
\begin{equation}\label{1.6}
\tilde{R}_{ij} -\frac{1}{2}\tilde{R}g_{ij}= \tilde{\Sigma}_{ij} + \frac{3}{64}(\bar{\psi}\gamma^5\gamma^\tau\psi)(\bar{\psi}\gamma^5\gamma_\tau\psi)g_{ij}
\end{equation}
where the purely metric energy density tensor is 
\begin{equation}\label{3.15}
\tilde{\Sigma}_{ij}:=\frac{i}{4}\left(\bar\psi\Gamma_{(i}\tilde{D}_{j)}\psi - \tilde{D}_{(j}\bar\psi\Gamma_{i)}\psi\right)
\end{equation}
and the Dirac field equations handled in a similar way yield the final expression
\begin{equation}\label{1.7}
i\Gamma^{h}\tilde{D}_{h}\psi
- \frac{3}{16}\left[(\bar{\psi}\psi)
+i(i\bar{\psi}\gamma^5\psi)\gamma^5\right]\psi - m\psi=0
\end{equation}
in terms of the $\tilde{D}_i$ which denotes the purely metric spinorial-covariant derivative.

This theory is general, but of course it is possible to apply it to some specific case: if the spinor is single-handed then it verifies relationships of the type $\gamma^5\psi=\pm\psi$ and the above system of field equations \eqref{1.6}, \eqref{3.15} and \eqref{1.7} becomes
\begin{equation}
\tilde{R}_{ij}=\frac{i}{4}\left(\bar\psi\Gamma_{(i}\tilde{D}_{j)}\psi- \tilde{D}_{(j}\bar\psi\Gamma_{i)}\psi\right)
\label{einstein}
\end{equation}
and
\begin{equation}
i\Gamma^{h}\tilde{D}_{h}\psi=0
\label{dirac}
\end{equation}
showing that for single-handed particle no torsion is present after all.
\section{Spin eigenstates in Weyl-Lewis-Papapetrou spacetime}
In the following, the single-handed spinor will be chosen to be left-handed (obviously, right-handed spinors would have the same dynamical properties), and therefore the spinor will be chosen to have the general form
\begin{eqnarray}
\psi=\left(\begin{tabular}{c}
$L$\\ $0$
\end{tabular}\right)
\end{eqnarray}
where $L$ is the left-handed spinor (the right-handed would have been the lower component).

The two components of the semi-spinor are the two opposite spin eigenstates, so that by writing the semi-spinor as
\begin{eqnarray}
L=\left(\begin{tabular}{c}
$a$\\ $b$
\end{tabular}\right)
\end{eqnarray}
the solution of the eigenstate equation $\frac{1}{2}\sigma_{3}L=\pm L$ assign to the upper component $a$ the meaning of the spin-up projection and to the lower component $b$ the meaning of the spin-down projection: eigenstates can be chosen by vanishing one of the projections, so that it is with no loss of generality that we can assume $b=0$ and the remaining degree of freedom will be described in general by $a=\Xi e^{i\beta}$ where $\Xi$ and $\beta$ are real functions.

The form of the spinor is therefore 
\begin{eqnarray}
\psi=\Xi e^{i\beta}\left(\begin{tabular}{c}
$1$\\ $0$\\ $0$\\ $0$
\end{tabular}\right)
\label{campo}
\end{eqnarray}
and from it we may compute the $5$ spinorial bilinear fields: it turns out that only the vector bilinear fields $\bar{\psi}\gamma^\mu\psi$ and $\bar{\psi}\gamma^\mu\gamma^5\psi$ are different from zero, and in particular they only have the temporal and one spatial component; what this means is that for single-handed spin eigenstates of spinors the spatial components of the bilinear vectors select a privileged direction in space, identifiable with the direction of propagation, and consequently it becomes possible without loss of generality to restrict the spacetime to that displaying axial symmetry.

Such a spacetime is long known to be the Weyl-Lewis-Papapetrou spacetime, in which the metric has line element
\begin{equation}\label{Mpol}
{ds}^{2}= -B^2 (r^2\,d\theta^2 +dr^2) -A^2 (-W\,dt + d\varphi)^2 + C^2\,dt^2
\end{equation} 
where all functions $A(r,\theta)$, $B(r,\theta)$, $C(r,\theta)$ and $W(r,\theta)$ depend on the $r$ and $\theta$ variables only.

The spinorial-connection coefficients are expressed as
\begin{subequations}
\begin{equation}
\Omega_{r}=\frac{AW_{ r}\gamma^3\gamma^4}{4C}+\frac{B_{ \theta}\gamma^1\gamma^2}{2 r B}
\label{oo1}
\end{equation}
\begin{equation}
\Omega_{\theta}=-\left(\frac{B + rB_{ r}}{2B}\right)\gamma^1\gamma^2 + \frac{AW_{ \theta}}{4C}\gamma^3\gamma^4
\label{oo2}
\end{equation}
\begin{equation}
\Omega_{\varphi} = \frac{A^{2}W_{ r}}{4BC}\gamma^1\gamma^4 - \frac{A_{ r}}{2B}\gamma^1\gamma^3 + \frac{A^{2}W_{ \theta}}{4 rBC}\gamma^2\gamma^4 - \frac{A_{ \theta}}{2 rB}\gamma^2\gamma^3
\label{oo3}
\end{equation}
\begin{eqnarray}
\nn
\Omega_{t} = \left(\frac{AW_{ r}}{4B} + \frac{WA_{ r}}{2B}\right)\gamma^1\gamma^3 + \left(\frac{AW_{ \theta}}{4 rB}+\frac{WA_{ \theta}}{2 rB}\right)\gamma^2\gamma^3\\
+\left(\frac{C_{ r}}{2B}-\frac{A^2WW_{ r}}{4BC}\right)\gamma^1\gamma^4
+\left(\frac{C_{ \theta}}{2 rB}-\frac{A^2WW_{ \theta}}{4 rBC} \right)\gamma^2\gamma^4
\label{oo4}
\end{eqnarray}
\label{def01}
\end{subequations}
with $\psi_{x^i}\neq0$ since for spinors a possible dependence of the azimuthal angle and the time may always be present.

With this form of the metric, we can evaluate the Einstein equations \eqref{einstein} getting 
\begin{subequations}\label{eqEinstein}
\begin{equation}\label{eqc11}
\begin{split}
\frac{A^2W^2_{ r}}{2C^2} + \frac{A_{ r}B_{ r}}{AB} + \frac{C_{ r}B_{ r}}{CB} - \frac{A_{ r r}}{A} - \frac{C_{ r r}}{C} - \frac{B_{ r r}}{B} + \frac{B^2_{ r}}{B^2} - \frac{B_{ r}}{ rB} -\frac{A_{ \theta}B_{ \theta}}{r^2AB} - \frac{C_{ \theta}B_{ \theta}}{r^2CB} \\
- \frac{B_{ \theta \theta}}{r^2B} + \frac{B^2_{ \theta}}{r^2B^2} =0
\end{split}
\end{equation}
\begin{equation}\label{eqc22}
\begin{split}
\frac{A^2W^2_{ \theta}}{2C^2} - \frac{ rC_{ r}}{C} - \frac{C_{ \theta \theta}}{C} - \frac{r^2B_{ r r}}{B} - \frac{ rB_{ r}}{B} - \frac{B_{ \theta \theta}}{B} -\frac{r^2B_{ r}C_{ r}}{BC}
+ \frac{B_{ \theta}C_{ \theta}}{BC} + \frac{r^2B^2_{ r}}{B^2}\\
 + \frac{B^2_{ \theta}}{B^2} - \frac{ rA_{ r}}{A} - \frac{A_{ \theta \theta}}{A} - \frac{r^2A_{ r}B_{ r}}{AB} + \frac{A_{ \theta}B_{ \theta}}{AB} =0
\end{split}
\end{equation}
\begin{equation}\label{eqc33}
\begin{split}
- \frac{A^4W^2_{ r}}{2B^2C^2} - \frac{A^4W^2_{ \theta}}{2r^2B^2C^2} - \frac{AA_{ r}C_{ r}}{B^2C} - \frac{AA_{ r r}}{B^2} - \frac{AA_{ r}}{ rB^2} - \frac{AA_{ \theta}C_{ \theta}}{r^2B^2C} - \frac{AA_{ \theta \theta}}{r^2B^2} + \frac{A\Xi^2\beta_{ \varphi}}{2} =0
\end{split}
\end{equation}
\begin{equation}\label{eqc44}
\begin{split}
- \frac{A^4W^2W^2_{ r}}{2B^2C^2} - \frac{A^4W^2W^2_{ \theta}}{2r^2B^2C^2} - \frac{AW^2A_{ r}C_{ r}}{B^2C} - \frac{AW^2A_{ r r}}{B^2} - \frac{3AWA_{ r}W_{ r}}{B^2} + \frac{A^2WW_{ r}C_{ r}}{B^2C} - \frac{A^2WW_{ r r}}{B^2}\\
- \frac{A^2W^2_{ r}}{2B^2} - \frac{AW^2A_{ r}}{ rB^2} - \frac{A^2WW_{ r}}{ rB^2} - \frac{AW^2A_{ \theta}C_{ \theta}}{r^2B^2C} - \frac{AW^2A_{ \theta \theta}}{r^2B^2} - \frac{3AWW_{ \theta}A_{ \theta}}{r^2B^2} + \frac{A^2WW_{ \theta}C_{ \theta}}{r^2B^2C} - \frac{A^2WW_{ \theta \theta}}{r^2B^2}\\
 - \frac{A^2W^2_{ \theta}}{2r^2B^2} + \frac{CC_{ r}A_{ r}}{AB^2} + \frac{CC_{ r r}}{B^2} + \frac{CC_{ r}}{ rB^2} + \frac{CC_{ \theta}A_{ \theta}}{r^2AB^2} + \frac{CC_{ \theta \theta}}{r^2B^2} - \frac{AW\Xi^2\beta_{ t}}{2} + \frac{C\Xi^2\beta_{ t}}{2} =0
\end{split}
\end{equation}
\begin{equation}\label{eqc12}
\frac{A^2W_{ r}W_{ \theta}}{2C^2} + \frac{A_{ \theta}B_{ r}}{AB} + \frac{A_{ r}B_{ \theta}}{AB} + \frac{B_{ r}C_{ \theta}}{CB} - \frac{A_{ r \theta}}{A} - \frac{C_{ r \theta}}{C} + \frac{C_{ r}B_{ \theta}}{CB} + \frac{A_{ \theta}}{ rA}  + \frac{C_{ \theta}}{ rC} =0
\end{equation}
\begin{equation}\label{eqc13}
- \frac{A^2W_{ \theta}\Xi^2}{16 rC} + \frac{AB_{ \theta}\Xi^2}{8 rB} - \frac{A_{ \theta}\Xi^2}{8 r} + \frac{A\Xi^2\beta_{ r}}{4} =0
\end{equation}
\begin{equation}\label{eqc14}
\frac{\Xi^2C\beta_{ r}}{4} - \frac{\Xi^2\left(4AW\beta_{ r} r - AW_{ \theta} -2WA_{ \theta} + 2C_{ \theta}\right)}{16 r} + \frac{A^2WW_{ \theta}\Xi^2}{16 rC} + \frac{\Xi^2\left(CB_{ \theta} - AWB_{ \theta}\right)}{8 rB} =0
\end{equation}
\begin{equation}\label{eqc23}
-\frac{ rAB_{ r}\Xi^2}{8B} + \frac{ rA_{ r}\Xi^2}{8} + \frac{ rA^2W_{ r}\Xi^2}{16C} + \frac{A\Xi^2\beta_{ \theta}}{4} - \frac{A\Xi^2}{8} =0
\end{equation}
\begin{equation}\label{eqc24}
\begin{split}
\frac{ rC_{ r}\Xi^2}{8} + \frac{\left(- rC\Xi^2 +  rA\Xi^2 W\right)B_{ r}}{8B} - \frac{ rWA_{ r}\Xi^2}{8} - \frac{\left( rAC\Xi^2 +  rA^2W\Xi^2\right)W_{ r}}{16C} - \frac{\Xi^2\left(-4\beta_{ \theta}+2\right)C}{16}\\
- \frac{\Xi^2\left(4W\beta_{ \theta}-2W\right)A}{16} =0
\end{split}
\end{equation}
\begin{equation}\label{eqc34}
\begin{split}
\frac{\Xi^2C\beta_{ \varphi}}{4} - \frac{AW\Xi^2\beta_{ \varphi}}{4} + \frac{A\Xi^2\beta_{ t}}{4} + \frac{A^4WW_{ r}^2}{2B^2C^2} + \frac{A^4WW_{ \theta}^2}{2r^2B^2C^2} + \frac{AWA_{ r}C_{ r}}{B^2C} + \frac{AWA_{ r r}}{B^2} + \frac{3AA_{ r}W_{ r}}{2B^2}\\
- \frac{A^2C_{ r}W_{ r}}{2B^2C} + \frac{A^2W_{ r r}}{2B^2} + \frac{AWA_{ r}}{ rB^2} + \frac{A^2W_{ r}}{2 rB^2} + \frac{AWA_{ \theta}C_{ \theta}}{r^2B^2C} + \frac{AWA_{ \theta \theta}}{r^2B^2} + \frac{3AA_{ \theta}W_{ \theta}}{2r^2B^2} - \frac{A^2C_{ \theta}W_{ \theta}}{2r^2B^2C} \\
+ \frac{A^2W_{ \theta \theta}}{2r^2B^2} =0
\end{split}
\end{equation}
\end{subequations}
while with the spinorial-connection it is possible to expand the spinorial derivative so that Dirac equations \eqref{dirac} become
\begin{subequations}\label{eqDirac}
\begin{equation}\label{ef3r}
\frac{\Xi_{ \varphi}}{A} - \frac{\Xi_{ t}}{C} - \frac{W\Xi_{ \varphi}}{C}=0
\end{equation}
\begin{equation}\label{ef3i}
\frac{\Xi\beta_{ \varphi}}{A} - \frac{\Xi\beta_{ t}}{C} - \frac{W\Xi\beta_{ \varphi}}{C} =0
\end{equation}
\begin{equation}\label{ef4r}
-\frac{\Xi\beta_{ \theta}}{ rB} + \frac{\Xi_{ r}}{B} + \frac{\Xi B_{ r}}{2B^2} + \frac{\Xi}{2 rB} + \frac{\Xi A_{ r}}{2AB} - \frac{AW_{ r}\Xi}{4BC} + \frac{\Xi C_{ r}}{2BC}=0 
\end{equation}
\begin{equation}\label{ef4i}
\frac{\Xi\beta_{ r}}{B} + \frac{\Xi_{ \theta}}{ rB} + \frac{\Xi\beta_{ \theta}}{2 rB^2} + \frac{\Xi A_{ \theta}}{2 rAB} - \frac{A\Xi W_{ \theta}}{4 rBC} + \frac{\Xi C_{ \theta}}{2 rBC}=0
\end{equation}
\end{subequations}
where the separation between real and imaginary parts has already been done.

The first four of \eqref{eqEinstein} are the diagonal part of Einstein equations and they are those expected to determine the dynamics, while the remaining equations represent the off-diagonal part and they are expected to furnish constraints on the various components: in fact, there are $6$ independent functions of the spacetime point and $14$ independent equations, so that it is clear that some constraints on the functions must eventually be found.

In order to solve equations \eqref{eqDirac} and \eqref{eqEinstein}, we start by deriving $\beta_{ r}$ and $\beta_{ \theta}$ from equations \eqref{eqc13} and \eqref{eqc23} getting
\begin{subequations}\label{diffbeta}
\begin{equation}\label{sbb1}
\beta_{ r} = \frac{AW_{ \theta}}{4 rC} + \frac{A_{ \theta}}{2 rA} - \frac{B_{ \theta}}{2 rB}
\end{equation}
\begin{equation}\label{sbb2}
\beta_{ \theta} = - \frac{ rAW_{ r}}{4C} - \frac{ rA_{ r}}{2A} + \frac{ rB_{ r}}{2B} + \frac{1}{2}
\end{equation}
\end{subequations}
which can be inserted into \eqref{ef4r} multiplied by $B\Xi$ to yield equation
\begin{equation}\label{ef4rr}
\frac{A_{ r}}{A} + \frac{\Xi_{ r}}{\Xi} + \frac{C_{ r}}{2C} =0
\end{equation}
from which we get the relation
\begin{equation}\label{solxit}
\Xi = \frac{F_1( \theta, \varphi, t)}{A\sqrt{C}}
\end{equation}
in terms of $F_1$ as an arbitrary function independent on $ r$; also, inserting \eqref{sbb1}, \eqref{sbb2} and \eqref{solxit} into \eqref{ef4i}, we obtain
\begin{equation}
\frac{\left(F_1\right)_{ \theta}}{ rBA\sqrt{C}} =0
\end{equation}
from which it follows that $F_1$ is only function of $ \varphi$ and $ t$: all in all, we obtain
\begin{equation}\label{solxi}
\Xi = \frac{F_1( \varphi, t)}{A\sqrt{C}}
\end{equation}
showing that in the spinor module the dependence on azimuth and time and that on radial coordinate and elevation angle are encoded within two factorized functions. Inserting all these results into equation \eqref{ef3r}, after some calculations, we get the relation
\begin{equation}\label{ef3ss}
\frac{C( r, \theta)}{A( r, \theta)} - W( r, \theta) = \frac{\left(F_1\right)_{ t}}{\left(F_1\right)_{ \varphi}}
\end{equation}
from which we have necessarily
\begin{subequations}
\begin{equation}\label{ref3}
\frac{C}{A}-W = \lambda
\end{equation}
\begin{equation}\label{lef3}
\frac{\left(F_1\right)_{ t}}{\left(F_1\right)_{ \varphi}} =\lambda
\end{equation}
\end{subequations}
for some constant $\lambda$; equation \eqref{lef3} implies $F_1\left( \varphi, t\right) =F_2\left( \varphi+\lambda  t\right)$ for some arbitrary function $F_2$ and thus
\begin{equation}\label{solc1}
C=AW+\lambda A
\end{equation}
so that
\begin{equation}\label{solxi2}
\Xi = \frac{F_2\left( \varphi + \lambda  t\right)}{A\sqrt{C}}
\end{equation}
in which the spinor module the dependence on azimuth and time has been fixed. Now, inserting the content of equations \eqref{sbb1}, \eqref{sbb2}, \eqref{solc1} and \eqref{solxi2} into \eqref{ef3i} we get the further equation
\begin{equation}\label{edi}
\frac{F_2\left( \varphi + \lambda  t\right)\left(\beta_{ \varphi}\lambda - \beta_{ t}\right)}{A\left(A\left(W+\lambda\right)\right)^{\frac{3}{2}}} =0
\end{equation}
which gives
\begin{equation}\label{solb4}
\beta_{ t} = \lambda\beta_{ \varphi}
\end{equation}
fixing the interdependence of azimuth and time also in the phase of the spinor field. Collecting all results obtained so far, we have that if \eqref{sbb1}, \eqref{sbb2}, \eqref{solc1}, \eqref{solxi2} and \eqref{solb4} hold then all Dirac equations \eqref{eqDirac} are satisfied automatically.

Moreover, Einstein equations \eqref{eqc13}, \eqref{eqc14}, \eqref{eqc23} and \eqref{eqc24} are satisfied automatically as well. In addition to this, substituting \eqref{sbb1}, \eqref{sbb2}, \eqref{solc1}, \eqref{solxi2} and \eqref{solb4} into the remaining Einstein equations, differentiating with respect to $ \varphi$ and multiplying all by $2AC/F_2$, we obtain a unique equation given by
\begin{equation}\label{d3e}
F_2\beta_{ \varphi \varphi} + 2\beta_{ \varphi}F'_{2} =0
\end{equation}
where $F'_2$ denotes the derivative of $F_2$: this equation is integrated as
\begin{equation}\label{hhb3}
\beta_{ \varphi} = \frac{F_3\left( r, \theta, t\right)}{F_2^2}
\end{equation}
with $F_3\left( r, \theta, t\right)$ being an arbitrary function. Inserting \eqref{hhb3} into Einstein equations and performing the derivative of the so obtained equations with respect to $ t$ we can deduce that $\left(F_3\right)_{ t}=0$ and thus
\begin{subequations}
\begin{eqnarray}
\label{solb3b}
\beta_{ \varphi}= \frac{F_3\left( r, \theta\right)}{F_2^2}\\
\label{solb4b}
\beta_{ t} = \frac{F_3\left( r, \theta\right)\lambda}{F_2^2}
\end{eqnarray}
\end{subequations}
and furthermore, we can subtract the derivative of \eqref{solb3b} with respect to $ r$ to the derivative of \eqref{sbb1} with respect to $ \varphi$ and the derivative of \eqref{solb3b} with respect to $ \theta$ to the derivative of \eqref{sbb2} with respect to $ \varphi$ obtaining (after using \eqref{solc1}) that $(F_3)_{ r}=(F_3)_{ \theta}=0$ necessarily, so that $F_3$ has to be constant and
\begin{subequations}
\begin{eqnarray}\label{solb3bb}
\beta_{ \varphi}= \frac{k}{F_2^2}\\
\label{solb4bb}
\beta_{ t} = \frac{k\lambda}{F_2^2}
\end{eqnarray}
\end{subequations}
for some constant $k$ and the arbitrary function $F_2\left( \varphi +\lambda  t\right)$ above. 

Integrating \eqref{solb3bb} and \eqref{solb4bb}, we have
\begin{equation}\label{solbb}
\beta\left( r, \theta, \varphi, t\right) = G\left( \varphi + \lambda  t\right) + F_4\left( r, \theta\right)
\end{equation}
where $F_4$ is an arbitrary function of $ r$ and $ \theta$ while $G$ is a function of one variable, whose first derivative satisfies
\begin{equation}\label{dg}
G'\left( \varphi + \lambda  t\right) = \frac{k}{(F_2\left( \varphi + \lambda  t\right))^2}
\end{equation}  
while inserting \eqref{solc1} and \eqref{solbb} into \eqref{sbb1} and \eqref{sbb2} we get the equations
\begin{equation}\label{sF41}
\left(F_4\right)_{ r} = \frac{W_{ \theta}}{\left(4W+4\lambda\right) r} + \frac{A_{ \theta}}{2 rA} - \frac{B_{ \theta}}{2 rB}
\end{equation}
\begin{equation}\label{sF42}
\left(F_4\right)_{ \theta} = - \frac{ rW_{ r}}{4W+4\lambda} - \frac{ rA_{ r}}{2A} + \frac{ rB_{ r}}{2B} + \frac{1}{2}
\end{equation}
and therefore after integration of these last three equations, the dependence of the phase $\beta$ will be explicited.

To simplify things, we may introduce the function $V\left( r, \theta\right)$ such that
\begin{equation}\label{ssf4}
F_4\left( r, \theta\right) := V_{ \theta}
\end{equation}
with
\begin{equation}\label{solA}
A := \frac{Be^{2 rV_{ r}}}{\sqrt{W+\lambda}}
\end{equation}
and such that equation \eqref{sF41} is automatically satisfied while equation \eqref{sF42} becomes
\begin{equation}\label{eqV}
V_{ r r} = -\frac{V_{ r}}{ r} + \frac{1}{2r^2} - \frac{V_{ \theta \theta}}{r^2}
\end{equation}
representing an equation for the unknown function $V\left( r, \theta\right)$. Solutions of equation \eqref{eqV} have the form
\begin{equation}\label{tsv}
V\left( r, \theta\right) := \frac{\left(\ln{ r}\right)^2}{4} + \xi\left( r, \theta\right)
\end{equation}
in terms of the function $\xi\left( r, \theta\right)$, and indeed inserting \eqref{tsv} into \eqref{eqV} we obtain the equation
\begin{equation}\label{eqlpxi}
\Delta\xi=\xi_{ r r}+\frac{\xi_{ r}}{ r}+\frac{\xi_{ \theta \theta}}{r^2}=0
\end{equation}
in terms of the Laplacian in the plane $ r, \theta$ of the function $\xi$ showing that the function $\xi\left( r, \theta\right)$ has to be harmonic in the plane of the $ r, \theta$ coordinates. Summarizing all the obtained results, we have that the functions
\begin{subequations}\label{ora3}
\begin{equation}\label{ora3Xi}
\Xi = \frac{F_2\left( \varphi + \lambda  t\right)
\left(W+\lambda\right)^{\frac{1}{4}}e^{-3 r\xi_{ r}}}{\left( rB\right)^{\frac{3}{2}}}
\end{equation}
\begin{equation}\label{ora3beta}
\beta = G\left( \varphi + \lambda  t\right) + \xi_{ \theta}
\end{equation}
\begin{equation}\label{ora3C}
C =  re^{2 r\xi_{ r}}B\sqrt{W+\lambda}
\end{equation}
\begin{equation}\label{ora3A}
A := \frac{ re^{2 r\xi_{ r}}B}{\sqrt{W+\lambda}}
\end{equation}
\begin{equation}\label{ora3F2}
F_2\left( \varphi + \lambda  t\right) = \sqrt{\frac{k}{G'\left( \varphi + \lambda  t\right)}}
\end{equation}
\end{subequations}
satisfy the Dirac equations \eqref{eqDirac} and the Einstein equations \eqref{eqc13}, \eqref{eqc14}, \eqref{eqc23} and \eqref{eqc24}. 

We now solve the remaining Einstein equations: making use of \eqref{tsv}, \eqref{eqlpxi} and \eqref{ora3}, from equations \eqref{eqc11}, \eqref{eqc12} and \eqref{eqc22} we can derive the second derivatives of $B$, namely
\begin{subequations}\label{solBd}
\begin{equation}\label{dB11}
\begin{split}
B_{ r r} = \frac{1}{4r^2B}\left(4\left( rB\xi_{ r \theta}\right)^2 + 8 rB\xi_{ \theta r \theta} -12\left(B\xi_{ \theta \theta}\right)^2 + 4 rBB_{ r}\xi_{ \theta \theta} - 4 rBB_{ \theta}\xi_{ r \theta} \right.\\
\left. + 5\left( rB_{ r}\right)^2 + 4B^2\xi_{ \theta \theta} - 2 rBB_{ r} + B^2 - 3B^2_{ \theta}\right)
\end{split}
\end{equation}
\begin{equation}\label{dB22}
\begin{split}
B_{ \theta \theta} = - \frac{1}{4B}\left(12\left( rB\xi_{ r \theta}\right)^2 + 8 rB^{2}\xi_{ \theta r \theta} - 4\left(B\xi_{ \theta \theta}\right)^2 - 4 rBB_{ r}\xi_{ \theta \theta} + 4 rBB_{ \theta}\xi_{ r \theta} \right.\\
\left. + 3\left( rB_{ r}\right)^2 - 4B^2\xi_{ \theta \theta} + 6 rBB_{ r} + 3B^2 - 5B^2_{ \theta}  \right)
\end{split}
\end{equation}
\begin{equation}\label{dB12}
B_{ r \theta} = \frac{4 rB^2\xi_{ r \theta}\xi_{ \theta \theta} + 2B^2\xi_{ \theta \theta \theta} + 2 rB_{ r}B_{ \theta} + BB_{ \theta}}{ rB}
\end{equation}
\end{subequations}
and thus inserting equations \eqref{solBd} into \eqref{eqc33}, we obtain the expression of the Laplacian
\begin{equation}\label{sdw}
\begin{split}
\Delta W=\frac{\left(- kW^3 - 3k\lambda\/W^2 - 3k\lambda^2W -k\lambda^3\right)e^{-8 r\xi_{ r}}}{r^4B^2\left(W+\lambda\right)}\\
+ \frac{1}{r^4B^2\left(W+\lambda\right)}\left[\left(3r^4B^2_{ r} + \left(-12r^3\xi_{ \theta \theta} + 6r^3\right)BB_{ r} \right.\right.\\
\left.\left. + 3\left( rB_{ \theta}\right)^2 + 12r^3BB_{ \theta}\xi_{ r \theta} + \left(12r^4\xi^2_{ r \theta} + 12 \left( r\xi_{ \theta \theta}\right)^2 - 12r^2\xi_{ \theta \theta} + 3r^2\right)B^2\right)W^2 \right.\\    
+ \left(6\lambda r^4B^2_{ r} + \left(\left(-24r^3\xi_{ \theta \theta} + 12r^3\right)\lambda - 2r^4W_{ r}\right)BB_{ r} + 6\lambda\left( rB_{ \theta}\right)^2 \right.\\
\left. +\left(24\lambda r^3\xi_{ r \theta} - 2r^2W_{ \theta}\right)BB_{ \theta} + \left(\left(24r^4\xi_{ r \theta}^2 + 24r^2\xi_{ \theta \theta}^2 -24r^2\xi_{ \theta \theta} + 6r^2\right)\lambda \right.\right.\\
\left.\left. + 4r^3W_{ r}\xi_{ \theta \theta} - 4r^3W_{ \theta}\xi_{ r \theta} -2r^3W_{ r}\right)B^2\right)W + 3\lambda^2r^4B_{ r}^2 \\
+ \left(\left( - 12r^3\xi_{ \theta \theta} + 6r^3\right)\lambda^2 - 2\lambda r^4W_{ r}\right)BB_{ r} + 3\lambda^2r^2B_{ \theta}^2 + \left(12\lambda^2r^3\xi_{ r \theta} - 2\lambda r^2W_{ \theta}\right)BB_{ \theta}\\
+\left(\left( 12r^4\xi_{ r \theta}^2 + 12r^2\xi_{ \theta \theta}^2 - 12r^2\xi_{ \theta \theta} + 3r^2\right)\lambda^2 + \left(4r^3W_{ r}\xi_{ \theta \theta} - 4r^3W_{ \theta}\xi_{ r \theta} -2r^3W_{ r}\right)\lambda\right.\\
\left.\left. +2r^4W_{ r}^2 + 2r^2W_{ \theta}^2\right)B^2\right]
\end{split}
\end{equation}
which is a rather heavy expression. Nevertheless, it can be used together with equations \eqref{solBd} into \eqref{eqc34} and \eqref{eqc44}, and all remaining Einstein equations reduce to the unique final equation
\begin{equation}\label{efin}
6\left( r\xi_{ r \theta}\right)^2 + 6\xi_{ \theta \theta}^2 - \frac{6 rB_{ r}\xi_{ \theta \theta}}{B} + \frac{6 rB_{ \theta}\xi_{ r \theta}}{B} + \frac{3\left( rB_{ r}\right)^2}{2B^2} - 6\xi_{ \theta \theta} + \frac{3 rB_{ r}}{B} + \frac{3}{2} + \frac{3B_{ \theta}^2}{2B^2} =0
\end{equation}
in terms of $B$ and $\xi$ alone.

This equation too seems rather unmanageable, but it can be rewritten in the form
\begin{equation}\label{efinxx}
\left(2 r\xi_{ r \theta} + \frac{B_{ \theta}}{B}\right)^2 + \left(\frac{ rB_{ r}}{B} -2\xi_{ \theta \theta} + 1 \right)^2 =0
\end{equation}
and thus splitting into
\begin{subequations}\label{efinxxbis}
\begin{equation}\label{efin1}
\frac{ rB_{ r}}{B} -2\xi_{ \theta \theta} + 1 =0
\end{equation}
\begin{equation}\label{efin2}
2 r\xi_{ r \theta} + \frac{B_{ \theta}}{B} =0
\end{equation}
\end{subequations}
which can both be easily integrated yielding
\begin{equation}\label{solB11}
B = \frac{e^{-2 r\xi_{ r}}q^2}{ r}
\end{equation}
where $q$ is a suitable integration constant. Now, if we set the function $W( r, \theta)$ of the form
\begin{equation}\label{sow}
W( r, \theta) := \frac{1}{P( r, \theta)} - \lambda
\end{equation}
in terms of another arbitrary function $P( r, \theta)$, and insert \eqref{solB11} as well as \eqref{sow} in the expression of the Laplacian \eqref{sdw}, after some calculations we obtain the Poisson equation
\begin{equation}\label{eqpoisPE}
\Delta P=\frac{kB^2}{q^8}
\end{equation}
for the unknown function $P( r, \theta)$.

And at this point, the complete integration of the coupled system of Dirac and Einstein equations is over.

For each solution $\xi$ of the Laplace equation \eqref{eqlpxi}, one gets the function $B$ from equation \eqref{solB11} which allows to determine the function $P( r, \theta)$ and thus the function $W( r, \theta)$ through the Poisson equation \eqref{eqpoisPE} written as
\begin{eqnarray}
 r( rP_{ r})_{ r}+P_{ \theta \theta}=\frac{kr^2B^2}{q^8}
\end{eqnarray}
for simplicity: inserting the so-found solutions into equations \eqref{ora3}, we finally obtain
\begin{subequations}\label{ora4}
\begin{equation}\label{ora4Xi}
\Xi = \frac{\sqrt{k}}{q^3\sqrt{G'\left( \varphi + \lambda  t\right)}}\frac{1}{P^{\frac{1}{4}}}
\end{equation}
\begin{equation}\label{ora4beta}
\beta = G\left( \varphi + \lambda  t\right) + \xi_{ \theta}
\end{equation}
\begin{equation}\label{oraA}
A = q^2\sqrt{P}
\end{equation}
\begin{equation}\label{ora4B}
B = \frac{e^{-2 r\xi_{ r}}q^2}{ r}
\end{equation}
\begin{equation}\label{ora4C}
C = \frac{q^2}{\sqrt{P}}
\end{equation}
\begin{equation}\label{ora4W}
W( r, \theta) = \frac{1}{P( r, \theta)} - \lambda
\end{equation}
\end{subequations}
in terms of the constants $k,q,\lambda$, the function $G$ and the harmonic function $\xi( r, \theta)$. The corresponding line element is
\begin{equation}\label{mora4}
\begin{split}
ds^2=q^4\left[(2\lambda-\lambda^{2}P)d t^2+2(1-\lambda P)d \varphi d t
-Pd \varphi^2-e^{-4 r\xi_{ r}}d \theta^2
-e^{-4 r\xi_{ r}}\frac{d r^2}{r^2}\right]
\end{split}
\end{equation}
and the spinor field is given by 
\begin{eqnarray}\label{bionda4}
\psi=\frac{\sqrt{k}}{q^3}
\frac{e^{iG\left( \varphi + \lambda  t\right)}}{\sqrt{G'\left( \varphi + \lambda  t\right)}}
\frac{e^{i\xi_{ \theta}}}{P^{\frac{1}{4}}}\left(\begin{tabular}{c}
$1$\\ $0$\\ $0$\\ $0$
\end{tabular}\right)
\end{eqnarray}
as the most general exact solution of the spin eigenstates single-handed spinor fields.

For the sake of completeness, we stress that the determinant of the metric is
\begin{equation}
\sqrt{|\mathrm{det}g|}=\frac{q^8}{ re^{4 r\xi_{ r}}}
\label{det}
\end{equation}
which will also be used in further applications.

An interesting feature of solutions \eqref{mora4} and \eqref{bionda4} is that despite both the Riemann tensor and the spinor field be non-zero nevertheless all Riemann invariants vanish and so the metric is that of a VSI-spacetime \cite{Carminati,Zakhary} and both spinor invariants $\overline{\psi}\psi$ and $i\overline{\psi}\gamma^5\psi$ vanish and so the spinor is a flag-dipole \cite{s-g,daRocha:2013qhu}; in addition to this, with another direct check it is possible to see that the momentum density vector defined by $V^{i}=\Xi^{-2}P^{-\frac{1}{2}} \overline{\psi}\Gamma^{i}\psi$ is a covariantly constant null vector ($V^{2}=0$ and $D_{i}V_{j}=0$), therefore the metric represents a PP-wave spacetime, according to the definition introduced in literature in \cite{Ehlers} (see also \cite{Stephani}). Moreover, the fact that all invariants vanish means they all are regular, and this implies that any singularity is removable with a suitable choice of coordinates.

In the following we are going to study a special case.
\subsection{The radial case}
A first result that is necessary to stress is a no-go result: the assumption $W=0$ amounts to impose non-rotational structure to the metric \eqref{Mpol}, and in such circumstance it is immediate to see that from \eqref{ora4W} it follows that $P$ is constant and then from \eqref{eqpoisPE} we have $k=0$ yielding $\Xi=0$ necessarily, showing that apart from the vanishing spinor, no non-trivial solution can exist. This result may appear little in importance, but it shows that we are on the right track: the spinor might well be single-handed and a spin eigenstate but still it has a non-trivial rotational structure that cannot be compatible with the rotational invariance imposed by the $W=0$ constraint; the coupled system of Einstein-Dirac field equations is telling in terms of a dynamical solution what was expected in intuitive terms.

However, this does not mean it is impossible to study the purely radial case. The purely radial case is obtained, firstly, by assuming the solution $\xi$ of the Laplace equation \eqref{eqlpxi} of the form in which no elevation angle is present, and secondly, by considering also the solution $P$ of the Poisson equation \eqref{eqpoisPE} to have no elevation angle: the first assumption gives the possibility to solve the Laplace equation \eqref{eqlpxi} in general for
\begin{equation}\label{xic1}
\xi=k_0+k_1\ln{r}
\end{equation}
for some constants $k_0$ and $k_1$; this solution together with equation \eqref{ora4B} gives
\begin{equation}\label{eqpoisPcr}
P_{ r r} + \frac{P_{ r}}{ r} + \frac{P_{ \theta \theta}}{r^2} = \frac{ke^{-4k_1}}{r^2q^4}
\end{equation}
and this allows us to implement also the second assumption getting
\begin{equation}\label{solpcr1}
P= \tau_0 + \frac{ke^{-4k_1}\left(\ln{ r}\right)^2}{2q^4} + \tau_1\ln{ r}
\end{equation} 
as particular solution, $\tau_0$ and $\tau_1$ being suitable integration constants. Inserting equations \eqref{xic1} and \eqref{solpcr1} into \eqref{ora4}, we have the expression of the spinor
\begin{eqnarray}
\psi=\frac{\sqrt{k}}{q^3}
\frac{e^{iG\left( \varphi + \lambda  t\right)}}{\sqrt{G'\left( \varphi + \lambda  t\right)}}
\left(\tau_0 + \frac{ke^{-4k_1}\left(\ln{ r}\right)^2}{2q^4} 
+ \tau_1\ln{ r}\right)^{-\frac{1}{4}}\left(\begin{tabular}{c}
$1$\\ $0$\\ $0$\\ $0$
\end{tabular}\right)
\end{eqnarray}
and the line element
\begin{equation}\label{mora4crm}
\begin{split}
ds^2 = -\left[q^4\left(\lambda^2\tau_1\ln{ r} + \lambda^2\tau_0 -2\lambda\right)+ \frac{k\left(\ln{ r}\right)^2\lambda^2e^{-4k_1}}{2}\right]d t^2\\
-\left[2q^4\left(\lambda\tau_1\ln{ r} + \lambda\tau_0 -1\right) + k\lambda\left(\ln{ r}\right)^2e^{-4k_1}\right]d \varphi d t\\
-\left(q^4\tau_1\ln{r}+q^4\tau_0+\frac{1}{2}k\left(\ln{ r}\right)^2e^{-4k_1}\right)d \varphi^2\\
-q^4e^{-4k_1}d \theta^2-q^4e^{-4k_1}\frac{d r^2}{r^2} 
\end{split}
\end{equation}
showing that in the purely radial case, the general solution we found can be fully explicited.

However, more information is found by choosing a very special tuning for the parameters.
\subsubsection{The case $\tau_0=\tau_1=k_1=0$, $G(x)=x$, $k>0$}
A particular instance of the purely radial case \eqref{xic1} and \eqref{solpcr1} consists in setting $\tau_0=\tau_1=k_1=0$, $G(x)=x$ and $k>0$, where the line element assumes the form
\begin{equation}\label{llc1}
ds^2 = \left[2q^4\lambda - \frac{k\left(\ln{ r}\right)^2\lambda^2}{2}\right]dt^2 + \left[2q^4 - k\lambda\left(\ln{ r}\right)^2\right]d \varphi d t 
- \frac{k\left(\ln{ r}\right)^2}{2}d \varphi^2 
- q^4d \theta^2- \frac{q^4}{r^2}d r^2 
\end{equation}
while the spinor is
\begin{eqnarray}
\psi=\frac{(2k)^{\frac{1}{4}}}{q^2}e^{i\left( \varphi + \lambda  t\right)}
\frac{1}{(\ln{ r})^{\frac{1}{2}}}\left(\begin{tabular}{c}
$1$\\ $0$\\ $0$\\ $0$
\end{tabular}\right)
\end{eqnarray}
in which we see that $\ln{ r}$ must be positive, and therefore $r>1$ in general.

This suggest that the singularity of the metric \eqref{llc1} for $ r=0$ must be apparent, that is depending on the chosen coordinates: indeed, by performing the coordinate transformation
\begin{equation}\label{cc}
R=:\ln{ r}
\end{equation}
the line element \eqref{llc1} assumes the expression
\begin{equation}\label{llc1bis}
ds^2 = \left[2q^4\lambda - \frac{1}{2}k\lambda^2R^2\right]dt^2 + \left[2q^4 - k\lambda R^2\right]d\varphi dt - \frac{1}{2}kR^2d\varphi^2- q^4d\theta^2 - q^4dR^2 
\end{equation}
which is regular everywhere, as it should have been expected from the fact that no singularity was present for the curvature invariants; on the other hand, the spinor field still has a singularity on the sphere of unitary radius $r=1$ or in the new coordinates for $R=0$ as clear from
\begin{eqnarray}
\psi=\frac{(2k)^{\frac{1}{4}}}{q^2}e^{i\left( \varphi + \lambda  t\right)}
\frac{1}{R^{\frac{1}{2}}}\left(\begin{tabular}{c}
$1$\\ $0$\\ $0$\\ $0$
\end{tabular}\right)
\end{eqnarray}
diverging in the origin, although no singularity is known to be present for the spinorial invariants. It may be possible to track the origin of this apparent singularity in the fact that for massless spinors it is impossible to boost in the rest frame, and all information about energy and momentum will always be frame-dependent necessarily.

The computation of the energy of the spinor is done by integrating the energy density over the volume measured by the determinant of the metric \eqref{det} and it eventually gives
\begin{eqnarray}
E=\int_{V}\psi^{\dagger}\psi\sqrt{|\mathrm{det}g|}dV
=4\pi\sqrt{2k}q^4\int_{[0,\infty[}\frac{dR}{Re^{R}}
\end{eqnarray}
where $V$ is the occupied volume: the integral gives no problem at infinity, but in the origin it goes as $\ln{\frac{1}{R}}$ and so with a positive divergence. This logarithmic divergence recalls the one for ultraviolet regimes in quantum field theory.

Such divergence in quantum field theory is treated by applying a cut-off and insisting that no calculations be done after that, because the theory beyond the cut-off is expected to be different, and in the present context it is possible to establish a physical interpretation for this situation: the physics beyond the cut-off is different because there the energy density is so large that gravitational effects can no longer be neglected; equivalently, quantum field theory ceases to be valid beyond the cut-off because linearity can no longer be used when gravitation is considered.
\section{Conclusion}
In this paper, we have considered the massless neutral spinor field in interaction with its own gravity: as for the resulting mutually coupled system of non-linear differential field equations, we have found all of the exact solutions.

The solutions we have found have been proven to be in the form of PP-waves for the metric tensor and to be a flag-dipole for the spinor field: the flag-dipole fields we have found are more specifically the Weyl fermions, which are fundamental as the basic chiral components of Dirac fermions; and the property of the metric of being a PP-wave, and therefore of describing a VSI-spacetime, implies that such a metric has the radiative corrections that vanish identically, and therefore this spacetime after quantization is nothing more than its classical counterpart \cite{Coley:2008th}.

Or in a more picturesque language, in this case quantum gravity is nothing more than classical gravity.

On the other hand the interest for Weyl fermions has very recently received a considerable revival because of the discovery of Weyl fermions inside a synthetic metallic crystal called tantalum arsenide, an experimental finding that can be used to create massless electrons with no backscattering, therefore improving the efficiency of electronic devices such as large-volume single-mode lasers or for general approaches to quantum computing \cite{science}.

Before applications, it is also from a fundamental point of view that Weyl fermions are important: for instance, it is still open the possibility that neutrinos may oscillate because of intrinsic dynamics \cite{Fabbri:2015jaa}; hence, the persistence of the masslessness hypothesis for neutrinos would make them a fundamental example of Weyl fermions.

If neutrinos were not Weyl fermions because massive in principle, then their masses would most probably assigned from Yukawa couplings in terms of the Higgs mechanism after symmetry breaking; but before such a symmetry breaking neutrinos as well as all other fermions are still massless: therefore in the earliest stages of cosmic evolution, where the universe is supposed to be in a configuration of unbroken electroweak symmetry with all massless fermions and dominant gravitational contributions, the particles would be modelled by our solution.

More in general, our solution is important also from a purely mathematical perspective: because a Dirac fermion is a system of two Weyl fermions of opposite helicities with a mixing of the two opposite-helicity states given by the mass term then a massive fermion can be thought as a system constituted by two massless fermions plus a sort of interaction described by the mass term; consequently, solutions for the massive fermion can be sought in the form of exact solutions for the massless fermion plus corrections in small mass regimes. In such a perturbative expansion the leading contribution would be the exact solution we have presented here.

The found solution is therefore important not only because of its technological applications (large-volume single-mode lasers and quantum computing \cite{science}), but also because for the foundations of physics (approaches to quantum gravity \cite{Coley:2008th}); and of course the solution is important mainly because of the obvious mathematical interest of having a non-linear system of differential field equations exactly solved.

We believe that these results are of high importance for mathematics, as well as for physics in general, consequently prompting further research following this direction.

Extensions are already under investigation.

\end{document}